\documentstyle[psfig]{l-aa}

\def\ergsec{\hbox{erg s$^{-1}$}}

\def\degmark{^\circ}
\def \rsun {\ifmmode$R$_{\odot}\else R$_{\odot}$\fi}
\def \nh {N${\rm _H}$}
\def \hcm {\hbox {\ifmmode $ atoms cm$^{-2}\else atoms cm$^{-2}$\fi}}
\def \src {Her\,X-1}
\def\approxgt{\mathrel{\hbox{\rlap{\lower.55ex \hbox {$\sim$}}
        \kern-.3em \raise.4ex \hbox{$>$}}}}
\def\approxlt{\mathrel{\hbox{\rlap{\lower.55ex \hbox {$\sim$}}
        \kern-.3em \raise.4ex \hbox{$<$}}}}

\newcommand {\einstein} {{\it Einstein}}

\newcommand{\SAX}{{\em BeppoSAX\/}}

\newcommand {\eg} {{\it e.g.}}

\newcommand {\degree} {$^{\circ}$}

\newcommand {\rchisq} {$\chi_{\nu} ^{2}$}

\begin{document}

\title{The \SAX\ LECS X-ray spectrum of Hercules X-1}

\author{T.~Oosterbroek, A.N.~Parmar,
D.D.E.~Martin, U.~Lammers}

\offprints{T. Oosterbroek (toosterb@astro. estec.esa.nl)}
\institute{Astrophysics Division, Space Science Department of ESA, 
ESTEC, P.O. Box 299, 2200 AG Noordwijk, The Netherlands}

\thesaurus{08.09.2 \src; 08.14.1; 08.16.7 \src; 13.25.5)}

\date{Submission date: March 1996; Received date; accepted date}

\maketitle 

\markboth{T. Oosterbroek et al.: The \SAX\ LECS
X-ray spectrum of Hercules X-1}{T. Oosterbroek et al.: The \SAX\ LECS
X-ray spectrum of Hercules X-1}
 
\keywords{accretion, accretion disks --  X-rays: binaries  --
individual: \src} \\  

\begin{abstract}
We present 0.1--10~keV observations of \src\
obtained with the Low-Energy Spectrometer Concentrator instrument
onboard the \SAX\ satellite during the main on-state of the 35~day cycle.
We confirm the existence of an intense 0.093~keV blackbody component and 
a broad emission feature at 0.94~keV.  
The pulse phase dependence of these components is similar,
suggesting a common origin. This is most likely fluorescent
excitation of moderately ionized ($\xi\sim$10--100) material 
located at the inner edge of the accretion disk.
\end{abstract}

\section{Introduction}

\src\ is an eclipsing binary X-ray pulsar with a pulse period of
$\sim$1.24s and an orbital period of 1.7~days (Tananbaum et al.\ \cite{t:72};
Giacconi et al. 1973). 
The source exhibits a 35~day X-ray intensity cycle comprising a~
$\sim$10 day duration main on-state and a fainter $\sim$5 day duration 
secondary on-state approximately half a cycle later. At other phases of
the 35~day cycle, \src\ is still visible at a low level (Jones \& Forman
1976). This modulation has been ascribed to a tilted precessing
accretion disk that periodically obscures the line
of sight to the neutron star (Gerend \& Boynton 1980).  
In addition, a regular pattern of X-ray 
intensity dips are observed at certain orbital phases. These may be
caused by obscuration from periodically released matter from the companion
star (Crosa \& Boynton 1980). 

The broad-band on-state X-ray spectrum of \src\ is known to be complex and
consists of at least the following components:
(1) a power-law with a photon index, $\alpha$, of $\sim$0.9 in the energy 
range $\sim$2--20~keV. (2) cyclotron absorption (Tr\"umper et al. 1978; 
Mihara et al. 1990) at energies $\approxgt 20$~keV.
(3) A broad Fe emission feature near 6.4 keV (Pravdo et al. 1977; 
Choi et al.\ \cite{c:94}). (4) A $\sim$0.1~keV blackbody
(Shulman et al. 1975; Catura \& Acton 1975;
McCray et al. 1982; see also Mavromatakis 1993; Vrtilek et al. 1994;
Choi et al. 1997) and (5) a broad emission feature between 
0.8--1.4~keV, which may be unresolved Fe~L shell emission (McCray et al. 1982,
see also Mihara \& Soong 1994).

Pulse-phase spectroscopy using the \einstein\ Objective
Grating (OGS) and Solid State Spectrometers (SSS) in the energy range
0.15--4.5~keV by McCray et al. (1982) reveals that the maximum 
intensities of the blackbody and power-law components are shifted by
240$\degmark$. It is likely that the blackbody component 
results from hard X-rays that are reprocessed in the inner accretion disk.
McCray et al. (1982) note that the 
maximum intensity of the unresolved 0.8--1.4~keV feature appears 
coincident with that of the blackbody. 

The Low-Energy Concentrator Spectrometer (LECS) on board the \SAX\
satellite 
is an imaging gas scintillation proportional counter sensitive in the 
0.1--10~keV energy range (Parmar et al. 1997).
The LECS energy resolution
is a factor $\sim$2.4 better than that of the ROSAT Position
Sensitive Proportional Counter (PSPC; 0.1--2.5~keV; Tr\"umper 1983)
and comparable to that of the Solid State Imaging Spectrometer (SIS) 
instrument at energies $\approxlt 0.5$~keV (Tanaka et al. 1994). 
In this paper we utilize the broad-band coverage, moderate spectral 
resolution and good sensitivity of the LECS to investigate pulse 
phase dependent changes in the overall \src\ spectrum, paying particular
attention to the low energy components. 

\section{Observations}

\SAX\ observed \src\ at a 35 day phase, $\Phi_{35}$, of 0.07-0.15
(using the ephemeris of Wilson et al. (1993)) between 1996 July 24
13:34 and July 27 13:13 (UTC).
This was close to the expected intensity maximum 
which occurs at $\Phi_{35}\approx$0.1.
The standard LECS extraction radius of 8$'$ was used.
Good data were selected from intervals when the minimum elevation
angle above the Earth's limb was $>$4\degree\ and when the instrument
configuration was nominal using the SAXLEDAS 1.4.0 data analysis package.
The spectral analysis was performed with the response matrix from
the 1996 December 31 release of SAXDAS.
Since the LECS was only operated during satellite
night-time, this results in a total exposure of 61~ks.
The observation included eclipsing and dipping intervals which were
excluded from further analysis leaving an exposure of 37~ks.
All spectra were rebinned to have $>$20
counts in each bin to allow the use of the $\chi^2$ statistic.

\subsection{Phase averaged spectrum}
\begin{table}
\begin{center}
\caption[]{Fit parameters for the phase-averaged spectrum}
\begin{tabular}{ll}
\hline
\noalign {\smallskip}
      \nh\ ($\times10^{20}$~atoms~cm$^{-2}$)& $0.33\pm0.04$ \\
      Blackbody temperature (keV)           & $0.093\pm^{0.001}_{0.002}$ \\
      Blackbody normalization$^{1}$         & $(1.24\pm0.02)\times10^{-3}$ \\ 
      $\alpha$ (power-law photon index)     & $0.738\pm0.005$           \\
      Power-law normalization$^{2}$         & $(3.94\pm0.02)\times10^{-2}$         
 \\
      Fe L line energy (keV)                & $0.937\pm^{0.006}_{0.012}$           \\
      Fe L line FWHM (keV)                  & $0.45 \pm ^{0.02}_{0.01}$           \\
      Fe L line normalization$^{3}$         & $(1.86 \pm ^{0.11}_{0.06})\times10^{-2}$     \\
      Fe K line energy (keV)                & $6.39\pm0.03$           \\
      Fe K line FWHM (keV)                  & $1.88\pm0.14$           \\
      Fe K line normalization$^{3}$         & $(4.63\pm0.35)\times10^{-3}$           
\\
      \rchisq\                              & 1.38\\
      dof                                   & 913\\
\hline
\label{tab:ave_fits}
\end{tabular}
\end{center}
$^{1}L_{39}/(d_{10})^{2}$ with $L_{39}$ the
bolometric source luminosity in 10$^{39}$ ergs s$^{-1}$ 
and $d_{10}$ the distance
in units of 10 kpc, assuming isotropic emission;
$^{2}$photons keV$^{-1}$ cm$^{-2}$ s$^{-1}$ at 1
keV;
$^{3}$total photons cm$^{-2}$ s$^{-1}$ in the line.
\end{table}

\begin{figure*}
\centerline{
   \hbox{\psfig{figure=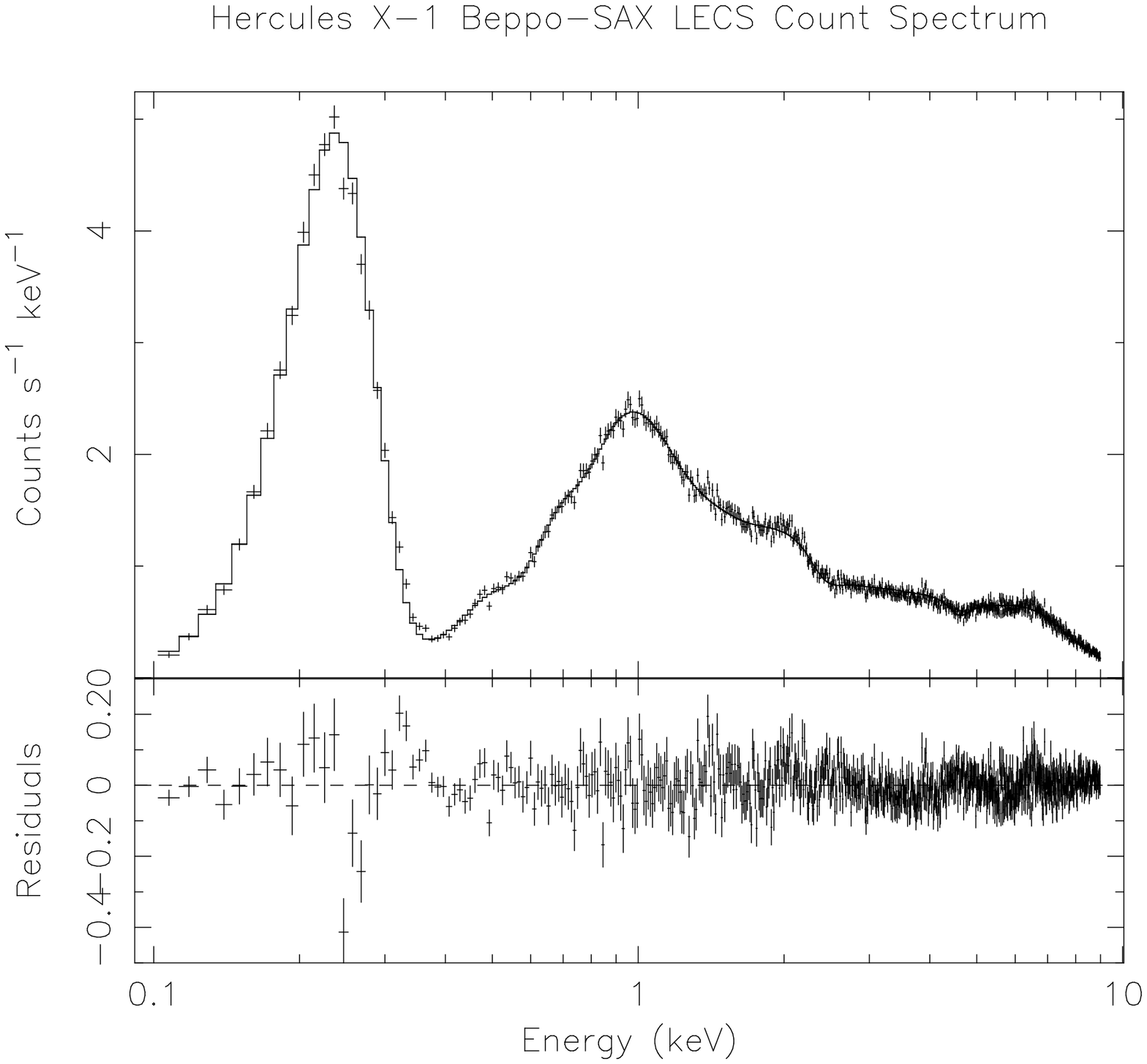,height=8.5cm,angle=-90,bbllx=50pt,bblly=75pt,bburx=575pt,bbury=650pt}
         \psfig{figure=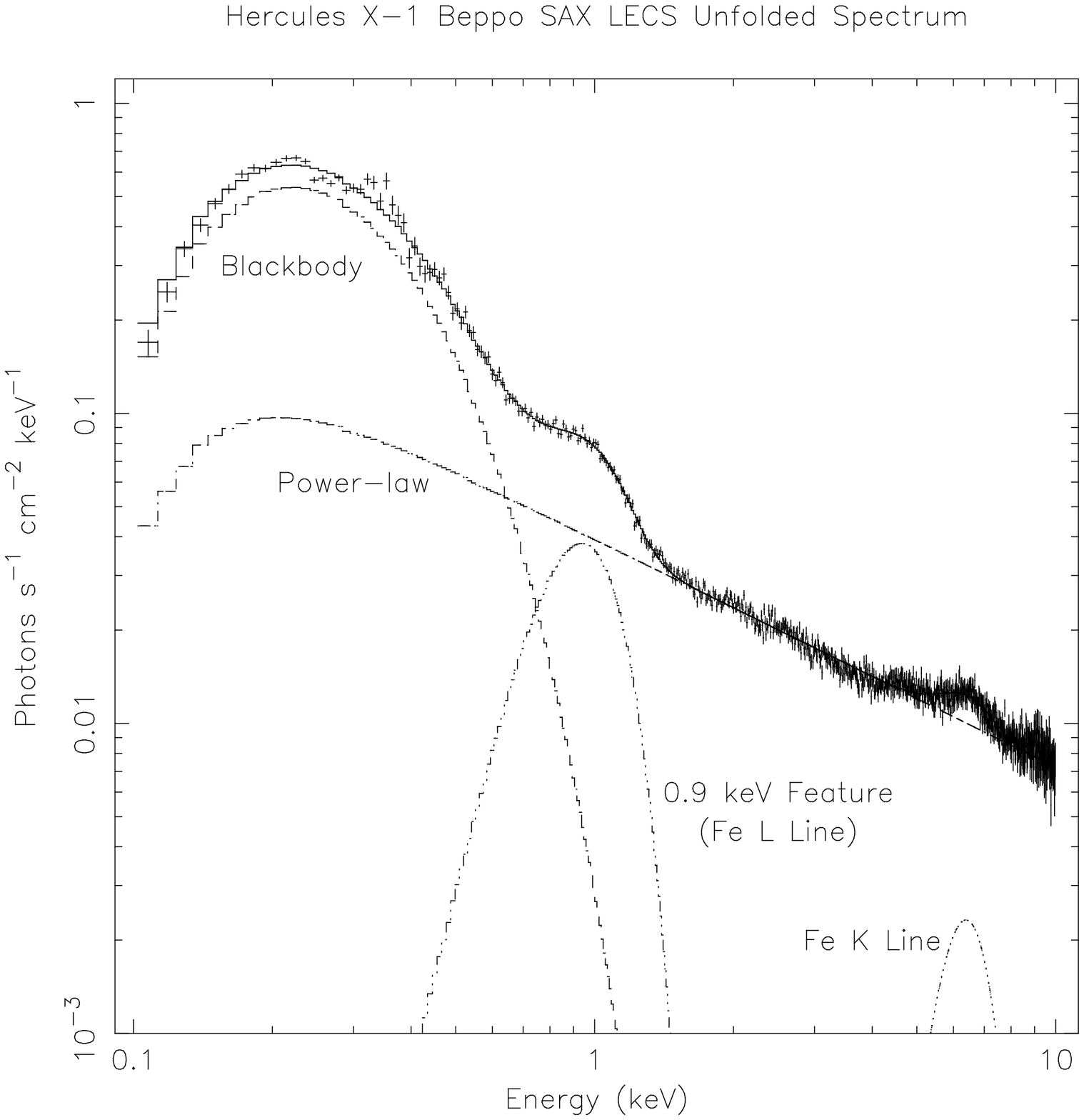,height=7.6cm,angle=-90,bbllx=72pt,bblly=64pt,bburx=600pt,bbury=550pt}}}
\caption[]{The LECS phase-averaged \src\ spectrum during dip and eclipse
free intervals. The lower left panel shows the residuals using the best-fit
model discussed in the text. The right panel shows the spectrum unfolded
using the best fit model and parameters given in the text. The individual
contributions are marked}
\label{fig:phase_ave}
\end{figure*}

We first investigated the phase averaged spectrum shown in 
Fig.~\ref{fig:phase_ave}. An intense soft flux from \src\ is clearly
indicated by the large number of counts visible 
below the instrument's entrance window cutoff at 0.28~keV.
For such a bright, high galactic latitude source, 
background subtraction is not critical and a standard 112~ksec
blank field exposure was used. 
A simple absorbed power-law model gives an unacceptable fit with
a \rchisq\ of 27 for 921 degrees of freedom (dof).
Adding a $\sim$0.1~keV blackbody improves the fit markedly, 
but still gives a \rchisq\ of 6.1. Including a broad 
Gaussian emission feature at 0.94~keV with an equivalent width (EW) of
410 eV reduces the \rchisq\
to 2.1. Finally, the addition of a Gaussian emission feature at 
6.4~keV with an EW of 460 eV 
gives a \rchisq\ of 1.38. Given 
the current calibration status of the LECS, it is 
not deemed worthwhile to add further components. 
Note that we cannot discriminate between a single broad Gaussian
line at 0.94~keV and two narrow lines at 0.90 and 1.06~keV 
seen in an extended low-state observation (Mihara \& Soong 1994).
The results of the phase averaged fit are given in Table~\ref{tab:ave_fits}.
All uncertainties are quoted at 68\% confidence.

The best-fit blackbody temperature of $0.093 \pm ^{0.001}_{0.002}$~keV
is consistent with that measured in the main on-state using the OGS
of $0.11 \pm 0.02$~keV by McCray et al. (1982). As in the OGS data, there is
no evidence for any emission features or edges directly associated 
with this component. The structure in the residuals 
(see Fig.~\ref{fig:phase_ave})
around 0.25~keV may indicate a more complex low-energy spectral
shape, or may result from small uncertainties in the LECS  
energy calibration. The soft component can also be
represented by a 0.34 keV bremsstrahlung model, which gives a similar value
of \rchisq, and similarly shaped residuals, but results in a value of \nh\ of
$(9.0 \pm 0.4)\times 10^{19}$~\hcm.
It is likely that the 0.94~keV feature seen in the LECS spectrum 
is the same component as seen in the SSS as an unresolved 0.8--1.4~keV feature 
and in the ASCA SIS as narrow 0.90 and 1.06~keV lines. This feature most
likely originates from Fe~L transitions and is hereafter referred to
as the ``Fe~L line''.
The blackbody to power-law (2--14 keV) luminosity ratio 
is $\sim$0.08, which differs slightly from
the value (0.13) obtained by McCray et al.\ (1982).

The best-fit value of the equivalent hydrogen column density, \nh, of
$(3.3 \pm 0.4)\times 10^{19}$~\hcm\ is significantly lower than that of
McCray et al. (1982) of 1.5--2.7$\times 10^{20}$~\hcm\ using the OGS
and that of Mavromatakis (1993) obtained during the secondary on-state 
with the PSPC and a power-law and blackbody spectral 
model of $(1.1 \pm ^{0.7}_{0.4})\times 10^{20}$~\hcm.
The \nh\  obtained with
a bremsstrahlung model is consistent with the value of Mavromatakis (1993).
The LECS \nh\ is consistent with 
the upper limit of $<3\times 10^{20}$~\hcm\ inferred from the absence
of a 2200\AA\ interstellar absorption feature in the spectrum of 
the counterpart star (Gursky et al. 1980).

\subsection{Pulse profile}
\begin{figure}
\centerline{\psfig{figure=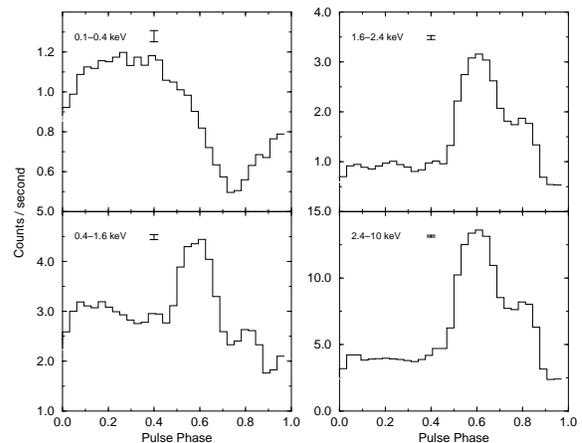,width=7.5cm,angle=-90,bbllx=63pt,bblly=37pt,bburx=594pt,bbury=710pt}}
\caption[]{\src\ pulse profiles during dip and eclipse free 
intervals in four energy bands. 
Pulse phase is arbitrary, 
$1\sigma$ uncertainties are indicated}
\label{fig:profile} 
\end{figure}
\begin{figure*}
\centerline{\psfig{figure=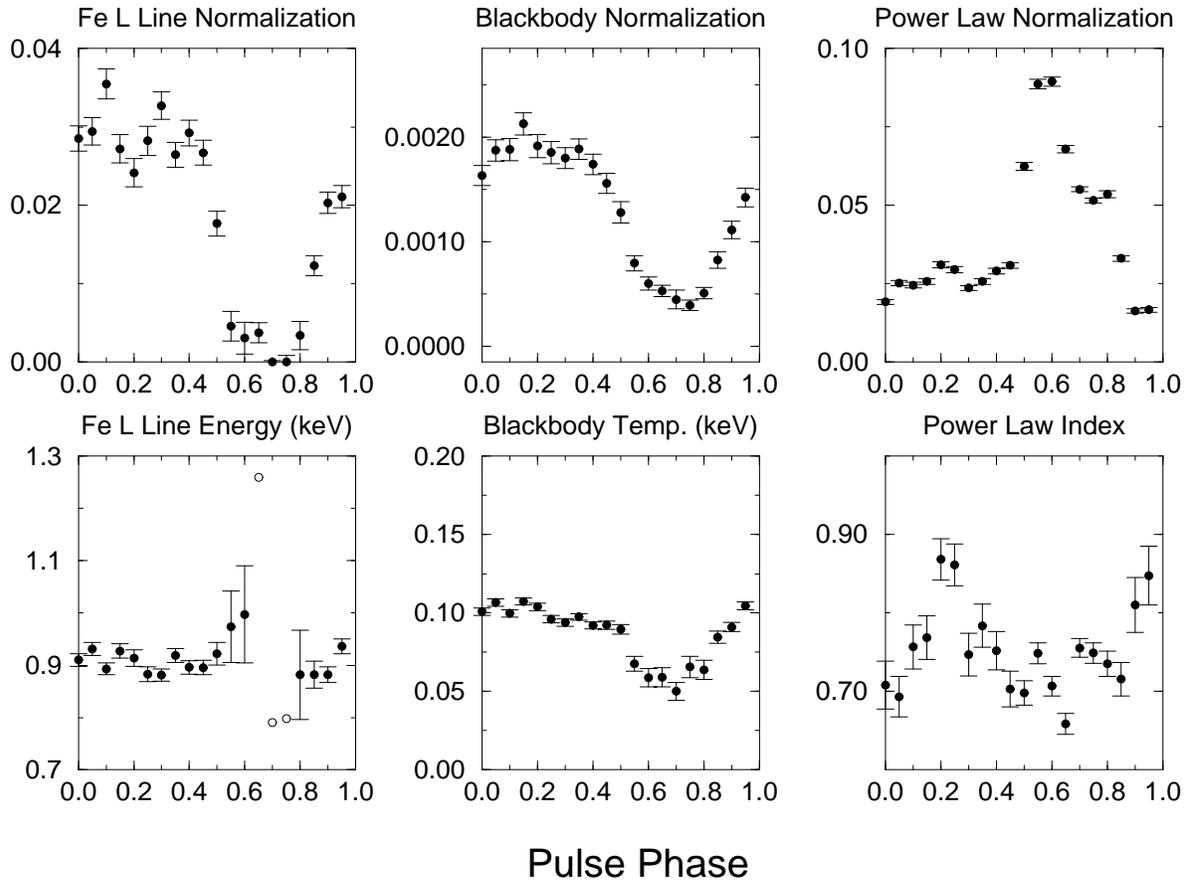,width=14cm,angle=-90,bbllx=50pt,bblly=50pt,bburx=535pt,bbury=650pt}}
\caption[]{Best-fit spectral parameters as a function of pulse
phase. In the lower left panel (Fe~L Line Energy) three points 
(plotted as open circles)
have large uncertainties, 
due to the low normalization of this component.
The units
of normalization are the same as used in Table~\ref{tab:ave_fits}}
\label{fig:phase_dep}
\end{figure*}
Event arrival times were first converted to the solar system barycenter.
Since the light travel time across the \src\ orbit ($\sim$26 s)
is large compared to the pulse period, the event
arrival times were additionally corrected to the \src\ center of mass 
using the ephemeris of Deeter et al.\ (\cite{d:91}). 
These corrected data were divided into 32 phase bins and
folded over a number of trial periods
and a pulse period of $1.2377396 \pm 0.0000001$~s obtained.
The uncertainty was determined by performing a linear least-squares
fit to the pulse arrival times.
This value is consistent with the pulse period history 
as obtained from BATSE observations.
The 0.1--10~keV pulse  
profile divided into four energy ranges is shown 
in Fig.\ \ref{fig:profile}.
This  confirms the phase shift between the hard 
($>$2 keV) and 
the soft ($<$0.4 keV) pulses first seen by McCray et al. (\cite{m:82}).

\subsection{Pulse phase resolved spectra}

We next investigated the pulse phase dependence of the various 
features identified
in the phase averaged spectrum. The corrected event data were folded 
into 20 phase bins using the pulse period given above.
The same model as used to describe the phase averaged spectrum was
fit to each of the spectra. Inspection of the results
revealed that the \nh, the central energy of the Fe~K line,
and the widths of the two
Fe lines did not vary significantly
with pulse phase, and so these were held fixed at the best-fit values given
in Table~\ref{tab:ave_fits}.

Figure~\ref{fig:phase_dep} shows the variations in the
best-fit spectral parameters as a function of pulse phase. 
Clear variations in all the parameters,
except for the energy of the Fe~L line, are visible. 
We note that the pulse phase dependence of the blackbody temperature
seems to mirror that of the power-law normalization. It is unclear
whether the observed change in the blackbody temperature results from
an actual change in the underlying blackbody, or from e.g.\
incorrect modeling of the power-law continuum. 
Any such effects are likely to be strongly phase dependent given the
large variations in normalizations of the two components (see Fig.\
\ref{fig:phase_dep}), and we note that the observed change in the
blackbody temperature occurs at pulse phases where the blackbody
component is faintest.

\section{Discussion}

There is a strong similarity between the pulse phase dependence of the
blackbody and 0.94~keV emission features (see Fig.~\ref{fig:phase_dep}).
This suggests that these spectral components
are physically related.
We have searched for any {\em orbital} phase dependence of this relation. 
Pulse phase resolved spectra were accumulated over 
orbital phases 0.06--0.45 and 0.45--0.72 (center of eclipse occurs at
phase 0.0). There is no significant difference in the behavior of the different
fit parameters in the two data sets. 

The LECS results are consistent with the illuminated inner part
of the accretion disk being the reprocessing site.
In a model where the accretion disk intercepts
a fraction of the hard X-ray beam (which is then reprocessed) a
180\degree\ phase difference between the hard X-rays and the reprocessed
X-rays is expected. If the disk is tilted with respect to the rotation 
axis of the
neutron star the phase difference can be different from 180\degree .
Since we find a $\sim$250\degree\ phase difference, this is consistent
with a ``tilted'' disk (for a more detailed discussion 
see McCray et al. \cite{m:82}). 
A way to test this is to observe \src\ at other $\Phi_{35}$.
If the tilt of the disk varies with $\Phi_{35}$, 
as predicted by the precessing disk models 
(\eg\ Gerend \& Boynton 1980), the phase difference should change.
Unfortunately, both the \SAX\ observation
and that of McCray et al. (1982) were performed at
approximately the same $\Phi_{35}$. We searched for
a change in the phase difference using data obtained during the
first and last 50~ksec of our observation (i.e.\ a total of $\sim$12~ksec 
exposure time).
We find a decrease of 11$\pm$50\degree\  which is inconclusive, mainly 
due to uncertainties in the determination of the peak of the broad
soft pulse.

Choi et al. (1994) find that the intensity of the 6.4~keV Fe~K 
line is modulated in phase with the ``soft'' component reported by
McCray et al. (1982). The LECS data is of insufficient quality to
investigate this relation, due to the low count rate and penetration
effects in the instrument (cf. Parmar et al. 1997). However, it appears
likely that both the Fe L and K lines exhibit a similar modulation,
and thus have the same origin. 

Kallman (1995) discusses Fe L/K
emission line ratios and mean line energies in photoionized gases. 
The line ratio depends on the shape of the ionizing continuum, 
the column density of the 
emitting material and the ionization parameter $\xi$ (Tarter et al.\ 1969).
The mean line energies increase with $\xi$, ranging from 
0.750 to 1.29~keV and 6.40 to 6.95~keV for Fe L and K, 
respectively.
Combining the calculations of Kallman (1995) with the
line energies and intensities given in Table~\ref{tab:ave_fits}
and assuming a power-law continuum with $\alpha$ in the range 0.5--1.0
allows $\xi$ and \nh\ to be constrained to be 10--100 and
$10^{22.5-23.5}$~\hcm,
respectively. The blackbody component could be
reprocessed emission originating 
at higher optical depths in the disk.
If the reprocessing region is located at the
magnetospheric radius of 3000~km, then the derived value of $\xi$ and
the peak power-law luminosity of $1.2\times10^{37}$ \ergsec gives an
\nh\ of only $10^{19.1}$--$10^{20.1}$ \hcm\ for a beam opening angle of
0.1 ($\Omega/4\pi$). However, it is likely that the reprocessing
region is not a single layer with uniform density, but more
complex, such that there are no unique values of \nh\ and $\xi$.

\acknowledgements 
T. Oosterbroek acknowledges an ESA Research Fellowship.
The \SAX\ satellite is a joint Italian and Dutch programme.

{}
\end{document}